\documentclass{icrc29}
\usepackage{graphicx,amssymb,amsmath,times}
\usepackage{picins}
\begin{document}
  \title[Spectral Monitoring of Mrk 421 during 2004]{Spectral Monitoring of Mrk 421 during 2004}
\author[J. Grube et al.] {J. Grube$^a$, for the VERITAS Collaboration$^b$ \\
  (a) School of Physics and Astronomy, University of Leeds, 
  Leeds LS2 9JT, United Kingdom \\
  (b) For full author list, see J. Holder's paper ``Status and Performance of the First VERITAS Telescope'' from these proceedings}
\presenter{Presenter: J. Grube (jg@ast.leeds.ac.uk), \ uki-grube-J-abs1-og23-oral}  \maketitle \begin{abstract} Daily monitoring of Mrk 421 in the X-ray and TeV bands during the first half 
of 2004 showed rapid flaring events above an elevated base flux level, 
culminating from March to May 2004 at an average flux level matching the highest 
detected flux state in 2001. Here 
we summarise the daily X-ray spectral variability observed with the RXTE-PCA 
instrument. We also explore the X-ray and TeV gamma-ray connection from near 
simultaneous data obtained using 
the Whipple 10m \v{C}erenkov imaging telescope. \end{abstract}  \section{Introduction} Mrk 421 is 
the archetype and closest known TeV blazar (z = 0.031), first 
detected at TeV $\gamma$-ray energies by the Whipple 10m telescope in 1992 
\cite{Punch92}. Blazars typically reside in large elliptical galaxies and 
are characterised by rapid variability at $\gamma$-ray energies, providing 
evidence of relativistic jet beaming along the line of sight \cite{Mara92}. 
Blazar surveys show a continuous sequence in Spectral Energy 
Distributions (SEDs) with TeV blazars at the high energy, low luminosity end 
\cite{Fossati98}. In the SSC model the first broadband peak in the SED 
(from UV to soft X-ray) is attributed to synchrotron radiation, and the second
 peak at GeV energies from inverse-Compton emission. 
Detailed studies of the X-ray spectrum over a broad energy range from 
0.1-25 keV detected by the satellite missions 
$BeppoSAX$, $ASCA$, $RXTE$, and $XMM$ find the synchrotron peak shifted to higher 
energies (0.5-2 keV) during high flux levels \cite{Xpeak}. Mrk 421 displays 
strong variability in the X-ray and $\gamma$-ray TeV bands on nearly all 
time-scales of hours to years with particularly strong flaring states 
detected by multiwavelength campaigns in 1995, 1998, 2001, 
and recently in 2004 \cite{everyone}. B{\l}a\.{z}ejowski et al. (2005) present 
a temporal analysis and correlation study of Whipple 10m $\gamma$-ray 
observations and near simultaneous $RXTE$ X-ray observations from December 2003 
to May 2004. This paper provides an extended analysis of the same data by 
resolving the X-ray spectrum on day time-scales and providing integral 
flux and hardness ratios for the 
TeV $\gamma$-ray data. \section{X-ray Observations and Data Analysis}
The NASA X-ray satellite $RXTE$ was launched in 1995, and is still in 
operation. The Proportional Counter Array (RXTE-PCA) instrument on board covers a
nominal energy range of 2-60 keV and consists of 5 PCU detectors. During the 
2004 campaign only PCU0 and PCU2 were operational. The daily ``snap-shot'' 
observations range from 0.6-16 ksec with a mean of 2.2 ksec. We followed the 
standard procedure for data reduction using FTOOLS 5.3 by filtering the 
data into Good Time Intervals and extracting a spectrum for individual 
observations. 
Using the latest background models and calibration 
tools simulated background spectra and response matrices were created 
with PCABACKEST and PCARSP.  For the high-state 
flux level during 12-23 April 2004 the spectra were fit over 28 bins from 4-15 keV.
All other spectra were rebinned by a factor of three due to low 
significances (4-6$\sigma$), resulting in nine bins covering 4-15 keV. 
Three fit 
models were tested: a simple power law, power law with an exponential cut-off, 
and a log-parabolic model. All fits included the fixed galactic column density 
$N_{H}=1.61 \cdot 10^{20}$ cm$^{-2}$ accounting for absorption from interstellar gas. 
Figure \ref{chi} shows the distribution of reduced $\chi^{2}$ goodness of fits 
for all observations to a power law and log-parabolic model.
It is clear a power law is not applicable to the data. 
The fits to an exponential cut-off give a marginally 
\piccaption{Reduced $\chi^{2}$ for power law fits (dashed) 
and log-parabolic fits (solid) to all RXTE-PCA spectra. \label {chi} }
\parpic[r][r]{\includegraphics*[width=0.33\textwidth,angle=0,clip]{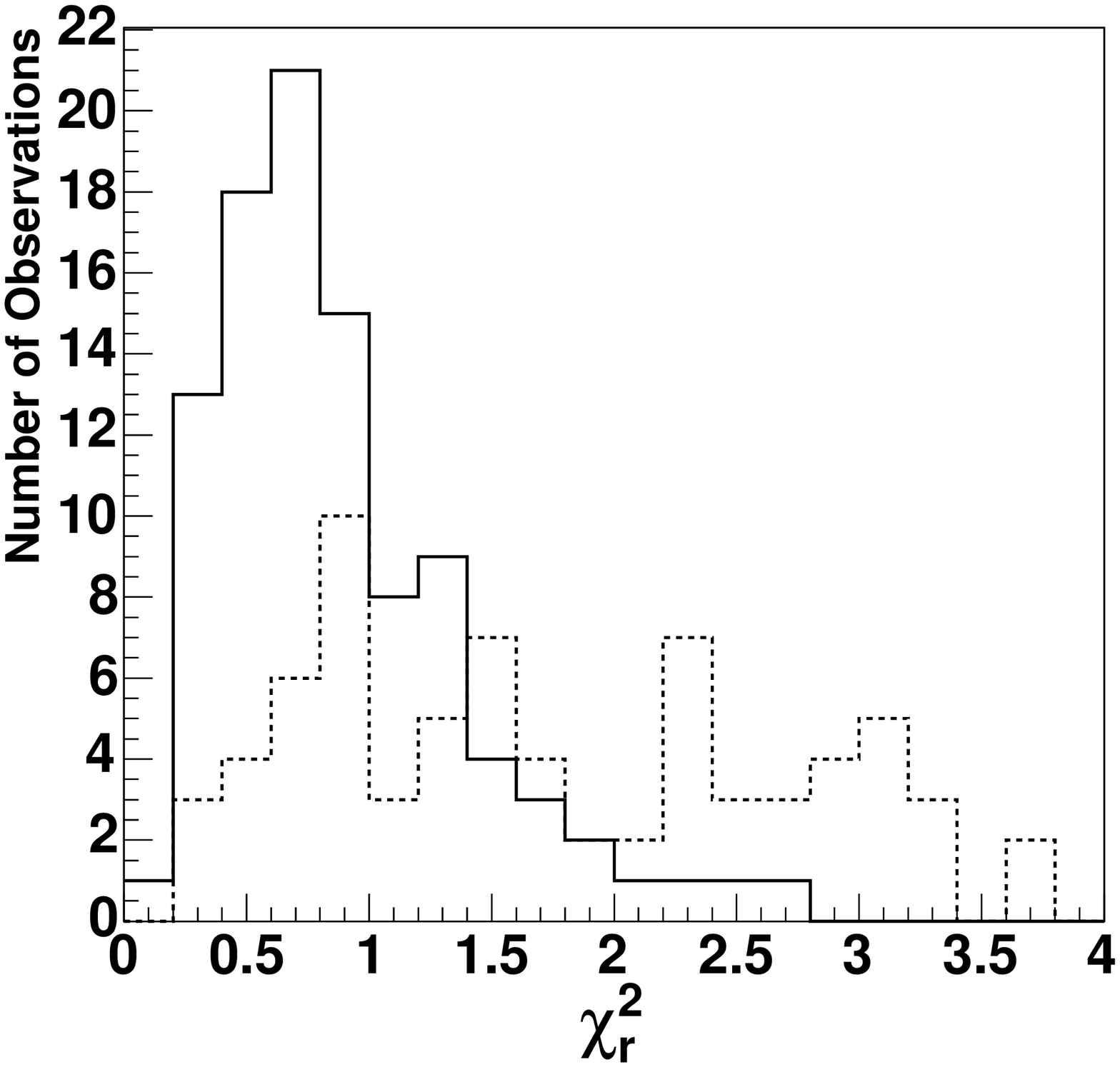}}
worse mean reduced $\chi^{2}$ of 0.98 compared to 0.87 for a log-parabolic 
model. Due to the limited energy coverage of 4-15 keV the calculated break 
energy in the exponential cut-off is unrealistically pushed to energies 
beyond the fit range ($E_{c}>$ 15 keV) \cite{Xpeak}. We therefore proceed 
with the log-parabolic fit described by 
\[F(E)=K(E/E_{1})^{-(a+b \cdot Log_{10}(E/E_{1}))}\] with $E_{1}$ set at 1 keV, 
giving three parameters: normalisation $K(E/E_{1})$, index $a$, and curvature 
term $b$ \cite{Xpeak}. The peak flux in the log-parabolic curve 
(synchrotron peak) is given as 
$\nu_{p}F(\nu_{p})=1.6 \cdot 10^{-9} \cdot K \cdot E_{1}^{2} \cdot 
10^{(2-a)^{2}/4b}$ in erg cm$^{-2}$ 
s$^{-1}$ and the peak energy as 
$E_{p}=E_{1} \cdot 10^{(2-a)/2b}$ in keV.  
\section{TeV $\gamma$-ray Observations and Data Analysis} \picskip{0}
The Whipple 10m telescope is located on Mt.~Hopkins, Arizona, USA 
at an elevation of 2312 m and is presently 
operating with the 379 pixel, $2.6^{\circ}$
field of view camera in place since 
1999. Monitoring Mrk 421 for high-state variability is a high 
priority; from November 2003 to May 2004 over 250 source $Tracking$ 
observations were carried out, each with a duration of 28 minutes. 
From this data set 141 runs met 
the criteria of zenith angle $< 28^{\circ}$ and relative throughput 
$> 0.7$ \cite{LeBohec03}. Monte Carlo 
simulations of $10^{6}$ $\gamma$-ray primary air showers with an energy 
distribution of $E^{-2.5}$ from 100 GeV to 50 TeV were generated with the 
GrISU package applying the 2004 telescope configuration for detection and 
analysis. The energy ``threshold'', defined as the position of the peak 
in the detected differential energy spectrum, is 
250 GeV at trigger level, and 400 GeV for events 
passing background rejection cuts described below.    

A combination of techniques from Whipple 10m, 
HEGRA CT1, and H.~E.~S.~S. data analysis are used here to derive integral 
flux values. After image cleaning, second moment and angle parameterisation, 
and trigger level cuts, 
$\gamma$/hadron ``extended'' cuts on $length$ and $width$ parameters were 
calculated from $\gamma$-ray simulations for a 90\% cut efficiency 
nearly independent of energy \cite{Mohanty98}. Due to the small 
field of view shower images at the edge of the camera are truncated. A 
cut on $distance$ of $< 0.9^{\circ}$ reduces this effect. The energy of 
each event is estimated using the method for HEGRA CT1 data described in 
\cite{Kranich01}. The $size$ parameter was scaled 
according to the throughput of each observation relative to a clear off field 
at $20^{\circ}$ zenith angle, providing a gain correction for nights with 
low atmospheric transparency \cite{LeBohec03}. A resolution in the shower core 
distance $I_{r}$ of $RMS(\Delta I)=$20\% and energy resolution $RMS(\Delta E)$ 
of 26\% for Whipple 10m simulations agree with that 
found for HEGRA CT1 \cite{Kranich01}. We apply a ``tracking'' analysis where 
the background region of the $alpha$ plot was used to estimate the background 
in the signal region $0^{\circ}<alpha<20^{\circ}$ \cite{everyone}.
An alternative approach is to match off fields from different observations to 
the $Tracking$ data based on zenith angle, throughput, and other factors. 
Comparing both techniques to a subset of Mrk 421 data with dedicated off field 
observations gives a systematic error of 6\% in the estimated background number, 
and 10\% in the integral flux value.

The integral flux is then simply the sum over the inverse effective areas 
calculated for each excess event, divided by the observation time. 
By using the 
``modified'' effective area (collection area as a function of estimated 
energy) the energy resolution and biases are implicitly accounted for 
\cite{Mohanty98}. The individual event effective areas are found by 
linear interpolation between logarithmic energy bins. The lower energy limit 
for integral flux values has been placed at 500 GeV, above which energy the 
effective area is relatively flat and independent of spectral shape. A hardness 
ratio was set for the bands I(0.5-1 TeV)/I(1-9 TeV). Data over the April 
2004 high-state flux level are resolved on nightly time-scales with exposures 
ranging from 0.5-2.8 hours. Data from lower flux levels are binned by 1-9 days, 
yielding higher statistics necessary in the high energy band. 
\section{Results and Conclusions} 
\begin{figure}[h]
\begin{center}
\includegraphics*[width=1\textwidth,angle=0,clip]{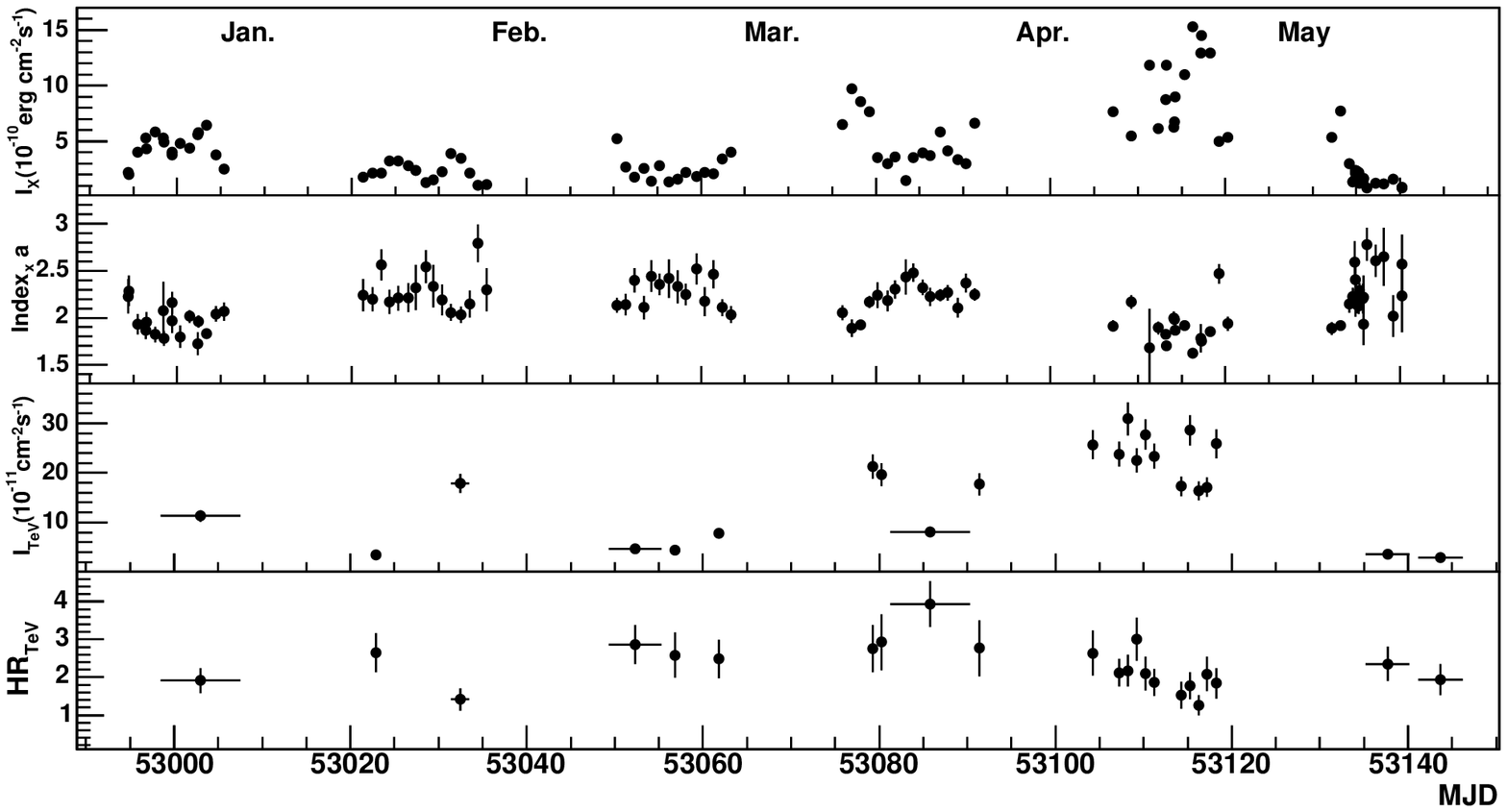}
\caption{Long-term light curve of Mrk 421 in 2004. The upper 
two panels show the RXTE-PCA 4-15 keV integral flux values, and spectral index 
``a'' from log-parabolic fits to the data. Below are Whipple 10m integral 
flux I(E$>$500 GeV) and hardness ratios I(0.5-1 TeV)/I(1-9 TeV) on nightly 
time-scales for the high-state phase in April 2004, and with varying exposures from 
1-9 days for the lower flux states. Systematic effects are included in the 
Whipple 10m flux errors.\label {Lcurve}}
\end{center}
\end{figure}
\begin{figure}
  \begin{center}
    \begin{minipage}[t]{0.49\linewidth}
      \includegraphics*[width=1.0\textwidth,angle=0,clip]{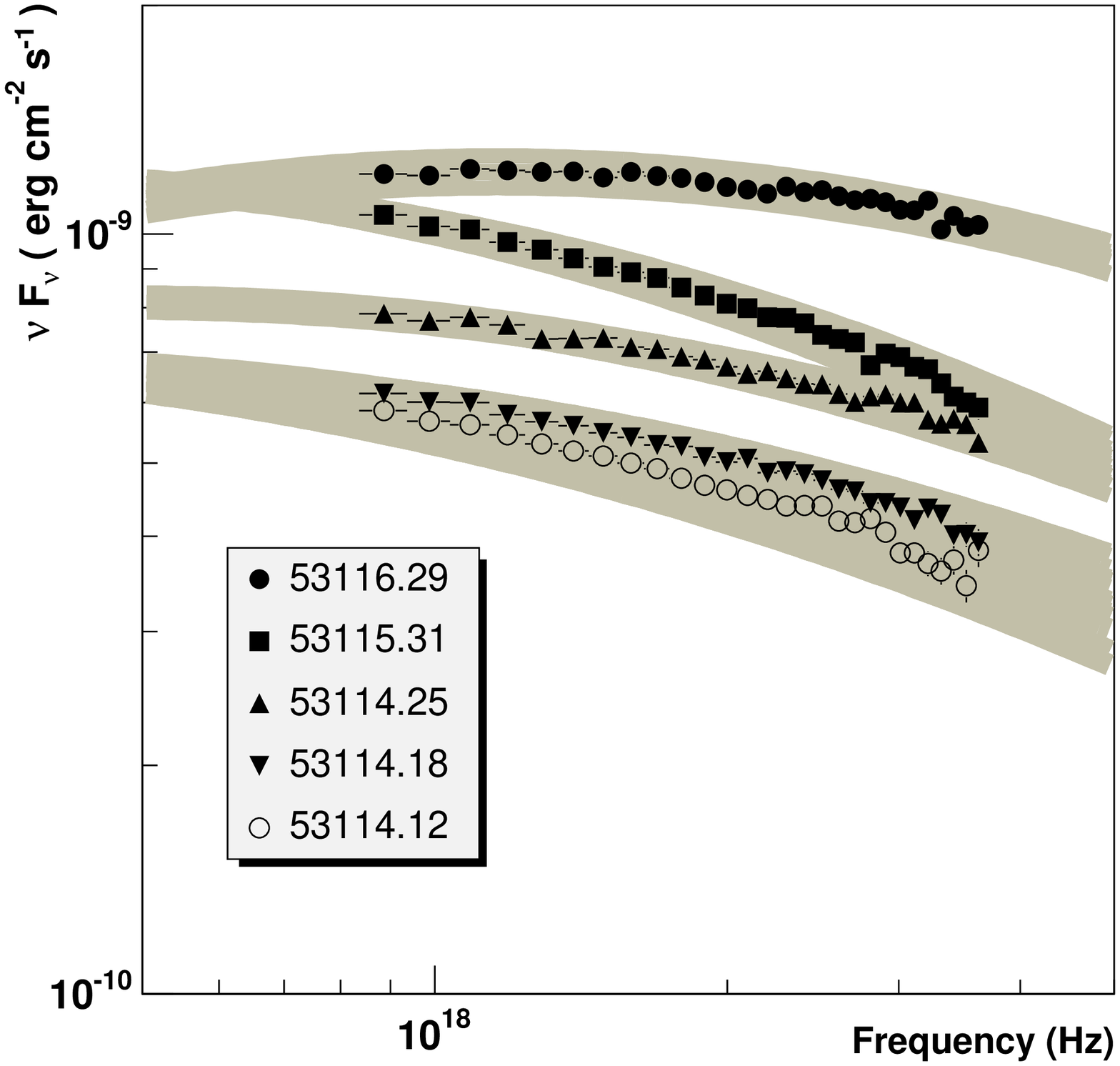}
      \caption{RXTE-PCA X-ray spectrum of Mrk 421 from 19-21 April 2004 
	(identified by MJD in plot). 
	The grey shaded bands show the 1 $\sigma$ statistical error region 
	from a log-parabolic fit. \label{SED}}
    \end{minipage}\hfill
    \begin{minipage}[t]{0.49\linewidth}
      \includegraphics*[width=1.0\textwidth,angle=0,clip]{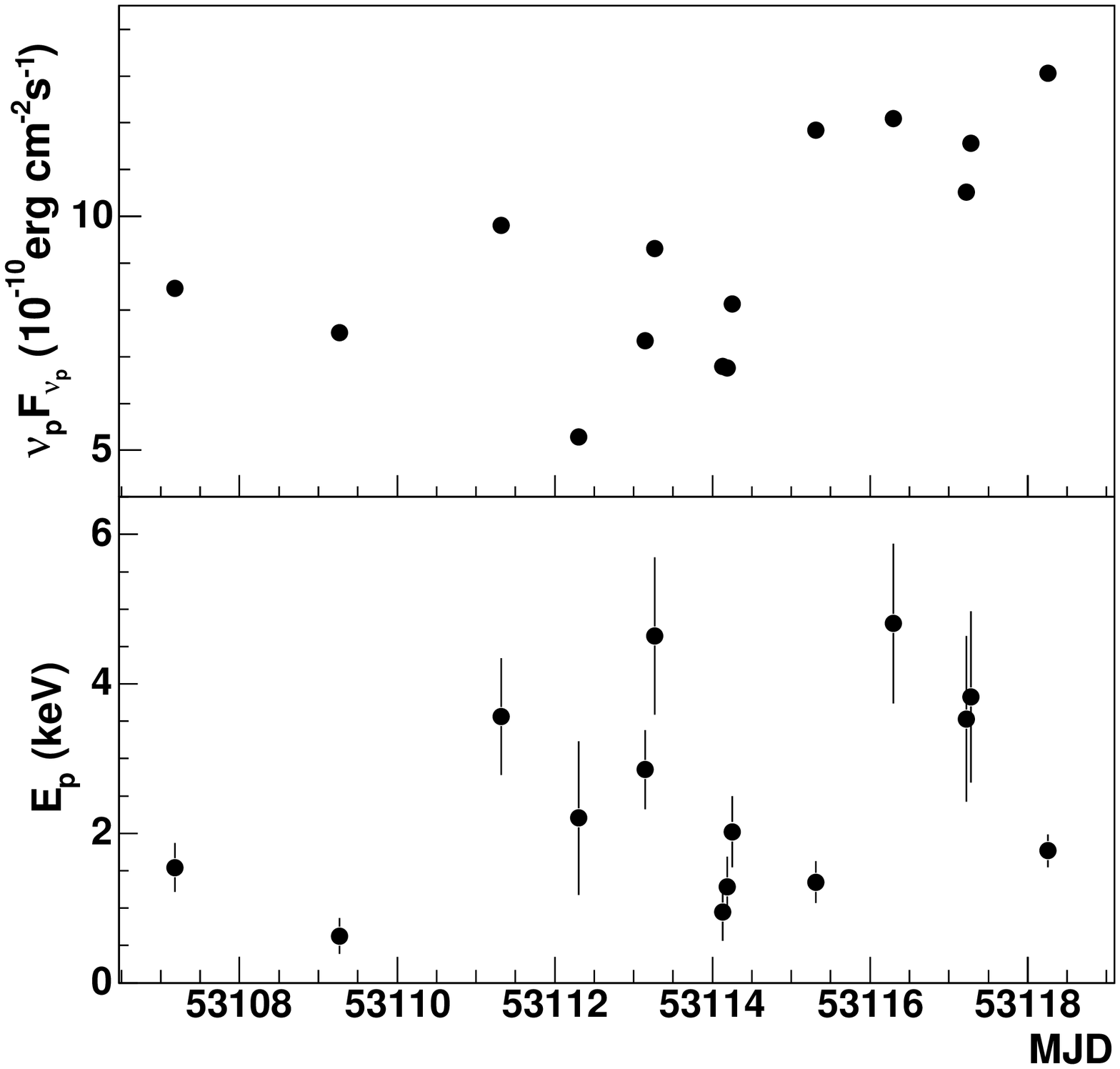}
      \caption{\label{Peak} Light curve of Mrk 421 peak synchrotron flux and energy 
	variability derived from log-parabolic fits to the RXTE-PCA X-ray spectrum 
	from 12-23 April 2004.}
    \end{minipage}
  \end{center}
\end{figure}

\piccaption{Correlation of RXTE-PCA spectral index ``a'' and integral flux over 
4-15 keV.\label {cor} }
\parpic[r][r]{\includegraphics*[width=0.33\textwidth,angle=0,clip]{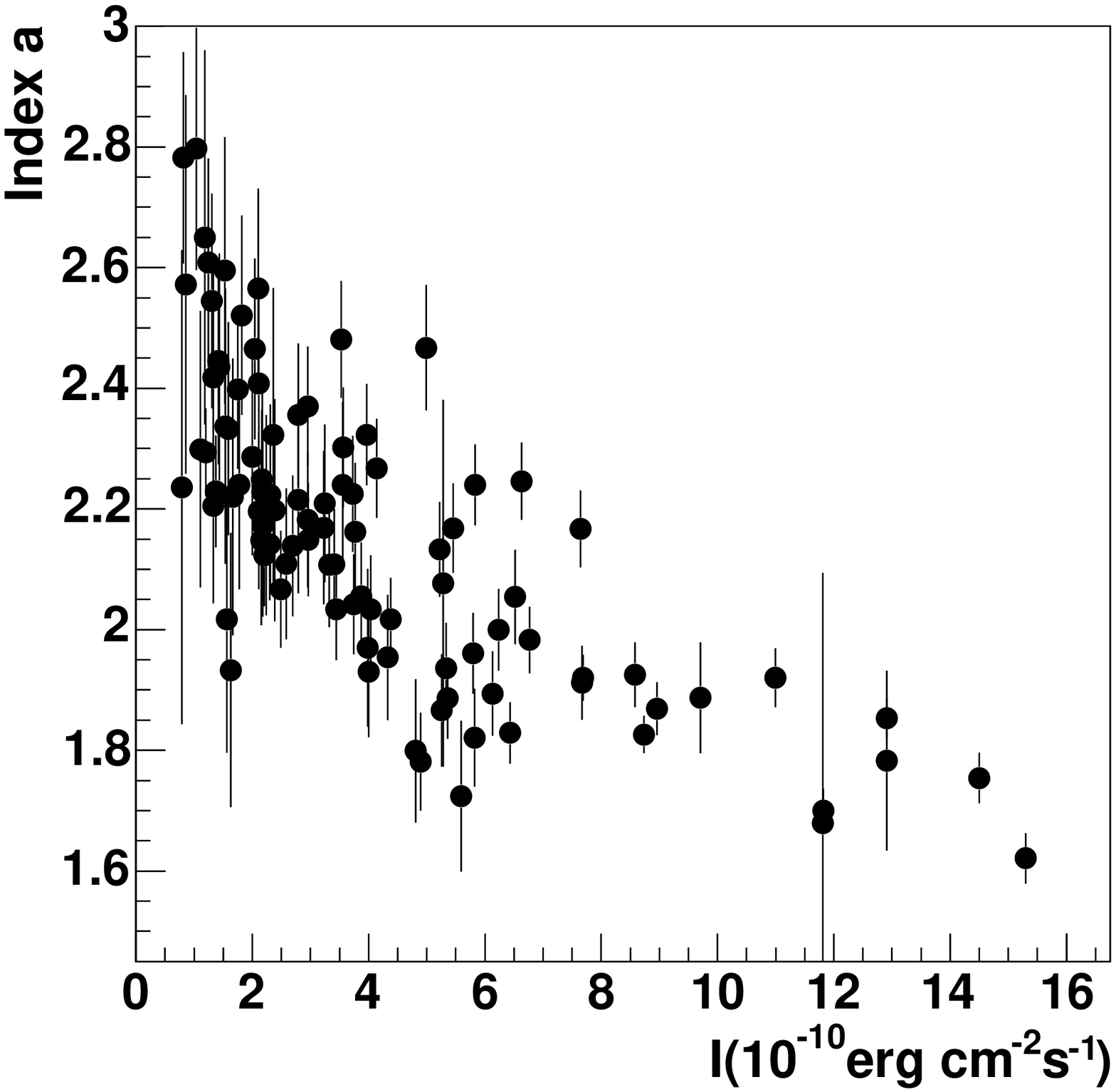}}
Figure \ref{Lcurve} shows the Mrk 421 light curve from December 2003 to May 2004. The 
RXTE-PCA X-ray integral flux I(4-15 keV) varies from a medium flux state to the 
near the highest yet detected, reaching the March 2001 peak value of 
$\sim$15 $\cdot $10$^{-10}$ erg cm$^{-2}$ s$^{-1}$ \cite{everyone}. 
The TeV $\gamma$-ray integral flux I(E$>$500 GeV) is variable over 
the range 0.28-2.8 $\cdot $10$^{-10}$ cm$^{-2}$ s$^{-1}$. The TeV $\gamma$-ray 
hardness ratios give an indication of spectral variability. 
As a reference, for the chosen intervals I(0.5-1 TeV)/I(1-9 TeV) spectra 
with power law indices ($\Gamma=$-2.0, -2.5, -3.0) yield hardness ratios 
(HR$=$1.1, 1.9, 3.0). The hardness ratios agree with previous results that 
find $\Gamma$ varying from -3 to -2 during 
medium to high flux states \cite{everyone}. Figure \ref{cor} shows the correlation 
of RXTE-PCA spectral index ``a'' to flux state. Previous X-ray spectral studies have 
shown rising flare events are initiated at high energies, resulting in a flatter 
spectrum at high flux levels \cite{Xpeak}. 
Figure \ref{SED} shows the rising RXTE-PCA spectrum in a high state during 
April 2004. The spectral shape stays nearly constant during the first day 
(53114 MJD). The spectrum for the following day\\ is steeper, 
with an increase in flux at lower energies. Finally, during the\\ detected flare peak 
on 53116 MJD the spectrum is very flat, evident \\of flaring at high energies. Due to 
the sparse data sampling the full evolution in the X-ray spectrum is obscured. 
Figure \ref{Peak} shows the calculated peak flux and peak energy E$_{p}$ values 
from the high-state X-ray spectrum during April 2004. Estimates of E$_{p}$ are limited 
by the fit range (4-15 keV). Still, we find the synchrotron peak shifted to the 
highest detected energies, reaching above 2 keV. Sampling synchrotron peak variability 
is necessary in deriving physical limits to emission mechanisms in TeV blazars. 
This campaign has allowed us to sample a wide range of flux states and flaring 
episodes, however future continuous multi-day X-ray observations of Mrk 421 with XMM 
and simultaneous TeV $\gamma$-ray observations with VERITAS promise a clear view of 
spectral variability on hourly flaring time-scales. 


\begin{thebibliography}{99}
\bibitem{Punch92} M. Punch et al., Nature 358, 477 (1992). 
\bibitem{Mara92} L. Maraschi, G. Ghisellini, and A. Celotti, ApJ 397, L5 (1992). 
\bibitem{Fossati98} G. Fossati et al., MNRAS 299, 433 (1998).
\bibitem{Xpeak}
E. Massaro, M. Perri, P. Giommi, and R. Nesci, A\&A 413, 489 (2004).

C. Tanihata, J. Kataoka, T. Takahashi, G. Madejski, ApJ 601, 759 (2004).

\bibitem{everyone}
J. Buckley et al., ApJ 472, L9 (1996)., L. Maraschi et al., ApJ 526, L81 (1999).

F. Aharonian et al., A\&A 410, 813 (2003)., M. B{\l}a\.{z}ejowski et al., ApJ accepted (2005).

\bibitem{LeBohec03}
S. LeBohec and J. Holder, Astropart. Phys. 19, 221 (2003).

\bibitem{Mohanty98}
G. Mohanty et al, Astropart. Phys. 9, 15 (1998).

\bibitem{Kranich01}
D. Kranich, PhD Thesis, Munich, Germany (2001).
\end{thebibliography}
\end{document}